# Effects of microstructure and growth conditions on quantum emitters in Gallium Nitride

Minh Nguyen,[1,†] Tongtong Zhu,[2,†] Mehran Kianinia,[1] Fabien Massabuau,[2] Igor Aharonovich,[1] Milos Toth,[1] Rachel Oliver,[2*] Carlo Bradac[1*]

[1]School of Mathematical and Physical Sciences, University of Technology Sydney, Ultimo, New South Wales, 2007, Australia
[2]Department of Materials Science and Metallurgy, University of Cambridge, Cambridge, CB3 0FS, United Kingdom

[†]These authors contributed equally to this work.
*Corresponding authors. Rachel Oliver: rao28@cam.ac.uk, Carlo Bradac: carlo.bradac@uts.edu.au

**Abstract**
Single-photon emitters in gallium nitride (GaN) are gaining interest as attractive quantum systems due to the well-established techniques for growth and nanofabrication of the host material, as well as its remarkable chemical stability and optoelectronic properties. We investigate the nature of such single-photon emitters in GaN with a systematic analysis of various samples produced under different growth conditions. We explore the effect that intrinsic structural defects (dislocations and stacking faults), doping and crystal orientation in GaN have on the formation of quantum emitters. We investigate the relationship between the position of the emitters—determined via spectroscopy and photoluminescence measurements—and the location of threading dislocations—characterised both via atomic force microscopy and cathodoluminescence. We find that quantum emitters do not correlate with stacking faults or dislocations; instead, they are more likely to originate from point defects or impurities whose density is modulated by the local extended defect density.

**Introduction**
Non-classical single photon emitters (SPEs) are one of the most fundamental components of quantum-based technologies—quantum computation,[1–3] quantum simulation,[4] quantum metrology[5] and quantum communication.[6] In particular, SPEs based on atom-like emitters in solid-state systems are widely studied. They combine the outstanding optical properties of atoms with the convenience and scalability of a solid-state host, hence constituting an ideal candidate for the practical realization of quantum-based devices. Some of the most prominent solid-state, room-temperature SPEs include colour centres in diamond,[7–9] silicon carbide,[10,11] zinc oxide[12,13] and hexagonal boron nitride.[14,15] In this context, single-photon emitters in gallium nitride (GaN) are emerging as an attractive alternative. The reason is twofold. The first is fundamental: recently GaN was found to host bright quantum emitters which are optically stable at room-temperature and emit over a wide range of wavelengths, from the UV to the infrared.[10,16,17] The second is practical: GaN is a well-known semiconductor with remarkable chemical stability and optoelectronic properties, in addition to established growth and nanofabrication techniques which have made it an ideal platform for integrated photonic, commercial LED and large-volume data storage technologies.[18,19] This means that GaN offers the unique opportunity—which is the focus of this study—to explore and understand the nature of the quantum emitters located within its crystalline matrix, while potentially allowing for the design of strategies for controlling and ultimately engineering their properties. This aspect is key, as controlling the quality—e.g. the accurate positioning of the emitters, the indistinguishability of the generated photons, as well as the ability to efficiently extract them from the crystalline matrix—is one of the major challenges many of the identified solid-state SPEs are still facing due to the complexity of the mesoscopic environment of the host material.[20]

GaN synthesis can be conducted via epitaxial growth on various substrates and following different protocols.[18,21] Currently, most growth methods still involve the use of foreign substrates such as silicon or sapphire. The flexibility to choose both procedures and substrate means that bulk GaN can be tailored to form distinct crystalline (hetero)structures such as a- and c-plane layers.[22–24] Growth of GaN on non-native substrates also means that the large lattice mismatch between the substrate and GaN can result in the formation of extended defects (e.g. threading dislocations (TDs)) and stacking faults, with the dominant defect depending on the growth orientation.[18,25] In addition, foreign atoms, e.g. Mg and Fe, can be readily introduced in the GaN matrix during synthesis.[26,27] Although GaN devices seem to be remarkably robust to defects compared to many other semiconductors, TDs do result in a deterioration of the optical and electronic performance observed in GaN devices[28,29]—e.g. through carrier scattering and non-radiative recombination processes. However, it is still unknown whether they play any role in the formation of quantum emitters. It has previously been hypothesised[10] that the formation of GaN quantum emitters is associated with extended structural defects created during GaN epitaxy. To test this hypothesis and explore the nature of the quantum emitters, we perform a systematic study of a number of GaN samples differing from one another in their density of TDs, the presence or absence of stacking faults, their doping, growth orientation, and geometry. We provide a detailed optical and spatial analysis of the quantum emitters found in each sample. We then investigate in detail the spatial correlation between the emitters found by photoluminescence (PL) and the position of the TDs in the GaN lattice identified by cathodoluminescence (CL) and atomic-force microscopy (AFM).

Quantum emitters in GaN have previously been hypothesised to relate to stacking faults.[10] We find no evidence for this contention in the current study. Instead, they are more likely to be due to point defects or impurities occurring at very low concentration but with a density which is affected by the local extended defect density. This work furthers our understanding of how quantum emitters are created in GaN and provides new information for the potential engineering of GaN-based scalable, integrated devices for advanced photonic and quantum applications.

**Results and Discussion**
*Sample growth and AFM characterisation:*
We analysed eight different GaN samples, labelled A–H, each one synthesised under specific growth conditions. The goal was to explore—and potentially tailor—the effects different parameters would have on the formation of quantum emitters in GaN. Each one of the eight samples was thus synthesised to control for a set of specific characteristics: density of dislocations (Samples A, B and C), doping of the matrix with extrinsic impurities expected to create deep-level states (e.g. with carbon (C) and iron (Fe) atoms: Samples D and E), growth substrate (e.g. GaN grown on silicon, Sample F) and orientation of the grown GaN material itself and hence the nature of the dominant defect (e.g. non-polar and semi-polar: Samples G and H). Samples A–E are c-plane layers in which the dominant defects are TDs, samples G and H are non- and semi-polar respectively and the dominant defects are basal plane stacking faults.

The samples were grown using metal-organic vapor phase epitaxy[18,29] on two different substrates, sapphire and silicon. Figures 1a–h show a schematic diagram of the cross section for each one of the different samples. Figures 1a'–h' show the corresponding topographic features—such as the surface terminations of crystallographic defects—as characterised by AFM. The following is a summary of the rationale behind the growth of each sample; the growth parameters are described in more detail in the Experimental section.

*i)* Samples A, B and C are c-plane (0001) GaN, which were deliberately grown under different conditions to have a substantial difference in the density of TDs in the GaN epitaxial layer: Sample A to have a high dislocation density of $\sim 5 \times 10^9$ cm$^{-2}$ (HDD) and Sample B a low dislocation density of $\sim 3 \times 10^8$ cm$^{-2}$ (LDD). An even lower density of TDs ($\sim 3 \times 10^7$ cm$^{-2}$ in this case) can be achieved using epitaxial lateral overgrowth (ELOG), Sample C. The intent for these samples is to establish whether the number and position of crystallographic defects correlate with the number and position of deep defect states in the material, which has been object of intense debate.[30]

*ii)* Samples D and E were grown with the intent to study the effects of extrinsic doping during growth, i.e. whether impurities which are expected to create deep-level states in the GaN bandgap would create optically-active colour centres within the GaN host matrix. Samples D and E were synthesised to intentionally contain a high concentration of Carbon (C-doped) at $1.5 \times 10^{19}$ cm$^{-3}$ and Iron impurities (Fe-doped) at $1 \times 10^{18}$ cm$^{-3}$, respectively.

*iii)* Sample F was synthesised to investigate the formation of emitters in relation to the substrate on which the GaN structure was grown on. In this respect, GaN is a favourable choice as the wide availability of different methods of GaN growth allows for its synthesis on a variety of different substrates. Specifically, Sample F consists of n-doped, c-plane GaN grown on silicon. This resulted in an estimated threading dislocation density of $2 \times 10^9$ cm$^{-2}$.

*iv)* Sample G and H were synthesised to explore the effect of crystal orientation on the potential creation of optically-active defects. Sample G is non-polar (11-20) GaN, while Sample H is semi-polar (11-22) GaN. Heteroepitaxial non- and semi-polar GaN are known to contain large densities of partial dislocations and stacking faults, but stacking faults may be considered the dominant defect. Sample G has an estimated TD density[31] of $3 \times 10^{10}$ cm$^{-2}$ and a basal plane stacking fault density of $3 \times 10^5$ cm$^{-1}$, while Sample H has an estimated TD density[32] of $4 \times 10^9$ cm$^{-2}$ and a basal plane stacking fault density of $2 \times 10^5$ cm$^{-1}$. Hence, we can use these samples to probe if stacking faults are playing any role in the formation of SPEs in GaN. Moreover, the unintentional impurity incorporation rate in semi-polar is faster than non-polar, so that the doping in the non- and semi-polar sample will also differ, with a much higher density of oxygen impurities in the semi-polar sample.[33]

*PL characterisation*

In order to probe quantum emission from SPEs in GaN at room temperature, we performed PL measurement on all the eight samples. The samples were characterized in a lab-built confocal microscope described elsewhere.[34] Briefly, the specimens were excited with a 532-nm continuous wave laser through a 0.9-NA air-objective. The emitted light was collected back through the same objective, spectrally filtered and sent either to a spectrometer or to two avalanche photodiodes (APDs) arranged in a Hanbury Brown and Twiss (HBT) interferometer configuration for single-photon detection.

The PL spectroscopy measurements carried out on the eight samples are summarized in Figure 2. All samples grown on sapphire substrates displayed an intense background PL emission peak at 695 nm—which is known to be due to chromium impurities in sapphire.[35] Samples A, B, E, and H showed strong PL emission at other wavelengths. Figure 2a, b, e, and h show the spectra collected from each sample—the analysed areas are displayed in the corresponding confocal scans (Fig. 2a', b', e', and h'). Note that due to the high background signal, the emitters could not be clearly identified in the PL confocal scan; we thus relied on spectroscopy measurements to isolate the individual emitters. Samples A (HDD) and B (LDD) showed a high density of emitters with zero phonon line (ZPL) wavelengths ranging from 570 nm to 750 nm. Sample E (Fe-doped) showed distinct PL emission peaks at ~710 nm and ~750 nm, but the density of emitters was qualitatively lower than that observed in Samples A and B. In samples C, D, F and G no emitters but only broad PL emission (Fig. 2c, d, f, and g) was observed (note: the sharp peak in 2f is the Raman peak from the silicon substrate).

To gather more information about the origin of the PL signal from the various samples, we conducted a more detailed analysis. For the HDD Sample A, LDD Sample B, Fe-doped Sample E and semi-polar Sample H we measured the depth along the z-axis (orthogonal to the surface) at which the emitters were located in the GaN crystal. This was achieved by collecting consecutive spectra of each emitter at multiple z-positions by varying the sample-to-objective distance in the confocal microscope. The plots from reference emitters in LDD, HDD, Fe-doped and semi-polar samples are shown in Figure 3a–d. Interestingly, we observed that the emission in the HDD, LDD and Fe-doped samples did not originate from emitters at the surface of the GaN crystal but rather from centres located deeper in the material, ~1–2 μm beneath the surface. Conversely, in the semi-polar sample H, the emitters were found to be mostly at the surface (see discussion below).

We also studied the photon statistics of the PL emission with a HBT interferometer to investigate the presence of single-photon sources. For the emitters in Sample A (HDD), we measured values for the second-order autocorrelation function $g^{(2)}(\tau=0) < 1$ (not background-corrected), yet not < 0.5—considered indicative of single-photon emission. This suggests that in this sample the centres are clustered in groups of at least a few emitters per confocal spot area (linear size ~500 nm). Single-photon emitters were instead found in Sample B (LDD). Figures 3e and f show typical PL spectra from isolated emitters in the LDD GaN sample. As stated above, the peak at 695 nm (Fig. 3e) originates from chromium impurities in the sapphire substrate and has been excluded in the autocorrelation measurement by applying a filter in the collection path—the spectral window analysed is shaded in red in Figures 3e and f. The relevant second-order autocorrelation measurements are displayed in Figures 3g and h which show values for $g^{(2)}(\tau=0)$ of 0.38 and 0.42, respectively (values are not background-corrected)—indicative of the quantum nature of the emitters.

Whilst the HDD and LDD samples were grown under the same conditions indicating that the nature of the emitters is likely the same, it is worth noting that the two samples involved different early stages of the growth, which results in different impurity (mainly oxygen) incorporation in the region adjacent to the GaN/sapphire interface. However, the depth of the emitters identified from the confocal PL measurements does not correspond to this near-interface region, and at the relevant depth we expect both samples to have similar (low) oxygen concentrations of $3 \times 10^{16}$ cm$^{-3}$. The two samples mainly differ in the density of TDs—these being higher in the HDD sample than in the LDD one—which seems to correlate with the corresponding density of emitters—again higher in the HDD sample than in the LDD one. To quantify this precisely, we analysed the confocal scans (5 × 5 µm² area) of the two samples and compared directly the total number of emitters. In the HDD Sample A, we identified 53 emitters versus the 17 emitters of the LDD Sample B. These amount to values for the planar density of emitters of $2.1 \times 10^8$ emitters/cm² and $0.7 \times 10^8$ emitters/cm² for samples A and B, respectively. The difference in the number of emitters between A and B can already be qualitatively seen in the confocal PL scans of Figures 3a' and b', where the amount of bright regions in Sample A appears much greater than that of Sample B. Simultaneously, the dislocation densities estimated from the growth process for the HDD and LDD GaN samples were ~5 × 10⁹ and ~3 × 10⁸ dislocations/cm², respectively. Our analysis suggests that a higher density of TDs corresponds to a higher density of emitters, with the caveats that the overall density of emitters is significantly lower (~4–25×) than that of the TDs, and that the TDs-to-emitters correlation is non-trivial—in our analysis a ~17× increase in TDs between Sample A and B corresponds to only a ~3× increase in emitter density. Note that for Sample A, due to the inability to identify single emitters, the total number of emitters is underestimated—in fact, each bright spot in the confocal scan contains a few emitters, as indicated by the value of the second order autocorrelation measurement: $0.5 < g^{(2)}(0) < 1$. Undoped LDD samples typically contain about 50% edge and 50% mixed TDs (of which 50% of the mixed dislocations are undissociated and 50% are dissociated—i.e. two partial dislocations connected by a nm-size stacking fault). However, undoped HDD samples typically contain about 80% edge and 20% mixed dislocations (of which 50% of the mixed dislocations are undissociated and 50% are dissociated).[36] Given that we do not see a correlation between emitters and the presence of stacking faults (see below), it is unlikely that mixed-type dissociated dislocations relate to emitters. Hence the density of emitters may correlate more closely with the density of mixed-type undissociated dislocations. We note that for the ELOG sample, which shows no emitters, the dislocation density varies periodically in bands across the surface and we might thus expect to see corresponding bands of emitters, but this effect is absent. Overall, our data lead to posit a link between dislocations and emitters—yet not a simple one-to-one correlation. It should be highlighted here that these data provide no evidence for the previously hypothesised link between emitters and stacking faults,[10] since there is no evidence for the presence of extended stacking faults in these samples, and the samples with a significant stacking fault density show few or no emitters. On the other hand, we did not find any emitters in Sample F that was grown on silicon, despite a similar dislocation density to Sample B. This again suggests the relationship between the emitters and TD density is non-trivial, and on a practical level suggests that sapphire may be the better substrate to use when trying to engineer emitters.

To further analyse the relationship between dislocations and emitters, we used a multi-microscopy approach (cf. Methods) to link the specific structural defects and optical signatures of those quantum emitters in GaN and carried out correlated characterisation of the GaN samples via CL and AFM and combined them with the PL data. The idea was to pinpoint the potential role of TDs on the formation of luminescent quantum emitters—i.e. establishing whether there is a direct spatial correlation between the positions of the emitters, characterised by PL, and the position of any TDs, measured by AFM and CL. Figure 4 summarizes the result for Sample B (LDD), chosen for reference. In the figure, the yellow box marks the same region analysed with the different techniques. Figure 4a is an AFM scan of the area: crystallographic defects terminating at the surface [blue circles] are visible as pits with a diameter of <10 nm. In this scan, mixed TDs are visible but the smaller pits related to edge TDs are not resolved. Figure 4b is the corresponding CL image: the dark spots arise due to non-radiative recombination at the dislocations sites[37] and relate to both mixed and edge TDs. Hence, where the CL image [black spots] and

AFM [blue dots] both show a feature in Figure 4c, it is likely to be a mixed dislocation, whereas when the CL image shows a feature which is not present in the AFM one, it is likely to be an edge dislocation. The density of TDs measured via CL is ~$4 \times 10^8$ cm$^{-2}$ and is consistent with the value of ~$3 \times 10^8$ cm$^{-2}$ estimated from the growth process. In Figure 4b the overlaid [red dots] indicate the positions of the emitters measured separately in PL. Figure 4c is an overlay of all these measurements and highlights a few interesting aspects.

We do not find any obvious correlation between the position of the emitters detected in PL ([red dots] in Fig. 4c) and the dislocations measured by AFM and CL ([blue dots] and [black spots] in Fig. 4c, respectively). We can thus exclude the possibility that TDs result in discrete states in the bandgap which localise the electron-hole pair. If they did, in Figure 4c every [blue dot] (AFM) would align with a [black spot] (CL) and a [red dot] (PL). Hence, we must consider why the TD density might correlate with the emitter density, without a direct spatial correlation between the two.

One possibility is that the sub-surface defect structure is not reflected in the CL and AFM measurements; we thus used transmission electron microscopy (TEM) to investigate the defect structure at the depth from which the emitters have been identified to originate in the confocal PL studies (1–2 μm below the top surface). Figure 5 shows the results of such TEM analysis for two different regions of the LDD sample, with Figures 5a and 5b ($g = 11\bar{2}0$) showing both edge and mixed dislocations, and Figure 5c and 5d ($g = 0002$) showing mixed and screw dislocations (no screw dislocation was observed in these images). In all cases the TDs run roughly vertically and at a depth of 1–2 μm below the surface; very little bending is seen, so that the CL and AFM images may be expected to be representative of the dislocation positions at this depth. At greater depths (3–5 μm below the surface, close to the GaN/sapphire interface) significant dislocation bending is observed, which relates to the 3D growth step used to reduce the dislocation density. However, the available data provide no evidence that dislocations—which lie within the c-plane of GaN, and do not thread—contribute to the formation of quantum emitters. We reiterate again here that the TEM data provide no evidence for the existence of stacking faults in this sample.

As noted above, whilst the most obvious difference between the LDD and HDD GaN is the dislocation density, the 3D growth mode used to generate LDD GaN does result in unintentional incorporation of oxygen in the near interface region. Whilst the average thickness of this unintentionally doped layer is only about 1 μm, this unintentional doping might extend further into the film, in places. For the HDD GaN, where no 3D growth step is used the unintentional doping throughout the film is low. For the ELOG sample, inclined facets which encourage the incorporation of unintentional dopants can persist for several microns, resulting in a layer of unintentionally n-doped material of significant thickness. Hence, samples with a higher incorporation of unintentional dopants, i.e. samples with a higher density of shallow donors, tend to have a lower density of quantum emitters. (The one exception to this is the semi-polar sample, in which some emitters are seen despite a high density of unintentional doping. However, this sample is altogether atypical since the emitters are observed at the surface in this case—while sub-surface for all the other samples—and these emitters may thus have a distinctly different origin). This point, coupled with the difference between Sample E (Fe-doped, where emitters are observed) and Sample D (C-doped, where emitters are not found) suggests that controlling the density and nature of impurities is crucial to engineering quantum emitters in GaN.

Given the lack of correlation between the actual dislocation position and the emitters, but the fact that there is a correlation between the emitters and the dislocation density, one must then consider whether extrinsic impurities or point defects might be the source of the emission and then ask how the density of such impurities away from the dislocations might be effected by the dislocation density. The following considerations hint at a potential interpretation. Local variations in dislocation density affect the strain state of the film, altering bulk and/or surface diffusion coefficients. Changes to the bulk diffusion coefficient could affect the diffusion of impurities from the sapphire into the GaN film, whilst changes to the surface or bulk diffusion coefficients may affect independently the density of non-impurity point defects. Processes occurring at surfaces will also be affected by changes to the surface morphology and we note that HDD GaN films are rougher, with more distorted step edges than LDD or ELOG films, potentially providing more sticking sites for impurity atoms and hence altering the incorporation rate of sparse impurities from the gas phase. (The low density of emitters indicates that such impurities must exist at such a low concentration that they could not be detected by secondary ion mass spectrometry or other typical compositional analysis techniques). Additionally, we remark that the structure of the emitters formed by the extrinsic defects could be complex—e.g. involving atoms and vacancies—and such that the defects could annihilate and/or not be optically active at the boundary regions formed by dislocations or the surface. All these considerations are consistent with our observation that the emitters seem to form separately from the dislocations and away from the surface.

To summarise, we carried out a systematic study of a number of GaN samples with different characteristics tailored through different synthesis conditions. The goal was to study the effect of certain specific parameters—density of the dislocations, doping with foreign atoms, structure and orientation of the grown GaN material—on the formation of single quantum emitters. We found no correlation between stacking faults and emitters; we found instead a rough—although not one-to-one—correlation between the number of emitters and the density of dislocations. However, the emitter locations and the dislocation sites are spatially distinct. The density of intentional and unintentional impurities is also observed to affect the quantum emitter density. Hence, we suggest that emitters relate to an as-yet-unidentified impurity or point defect in GaN whose density is affected by the extended defect density. Future engineering of single photon sources based on GaN will thus require engineering of both the impurity and extended defect density of these materials. Nevertheless the application of expensive bulk substrates, which tend to have low extended defect densities and a high density of unintentional n-dopants is likely to be unnecessary. Instead efforts to engineer these devices can focus on widely available GaN epitaxial layers on sapphire.

**Methods**
*Sample Fabrication*
Samples A and B consist of 2- and 4-µm thick undoped polar c-plane (0001) GaN grown on c-plane sapphire with different dislocation densities at $5 \times 10^9\,\mathrm{cm}^{-2}$ (high dislocation density, HDD) and $3 \times 10^8\,\mathrm{cm}^{-2}$ (low dislocation density, LDD), respectively. The HDD Sample A used a 2D growth method, where a 30-nm GaN nucleation layer (NL) was grown at 540 °C and followed by the GaN growth at a V-III ratio of 1310. In the LDD Sample B, the NL was annealed at 1020 °C and followed by 3D growth using a low V-III ratio (715) and island coalescence was then enhanced by increasing the V-III ratio to 1075.[38] The dislocation reduction was achieved as the dislocations bend over and annihilate one another during the 3D-2D coalescence process.[39]

Sample C is c-plane GaN grown using ex-situ, pre-patterned GaN/sapphire pseudo-substrates/seed layer (with a dislocation density of $5 \times 10^9$ cm$^{-2}$) with periodic SiO$_2$ stripes and epitaxial lateral overgrowth technique. The average dislocation density in the ELOG sample is $3 \times 10^7$ cm$^{-2}$. The ELOG process reduces the number of dislocations by physically stopping their propagation through the GaN film. The selected area epitaxy allows GaN to grow only where the underlying GaN seed layer is exposed (window region), whilst the GaN grows laterally over the mask (wing region), where dislocations would bend over during coalescence. The dislocation densities are measured to be $5 \times 10^7$ cm$^{-2}$ and $9 \times 10^6$ cm$^{-2}$ in the window and wing regions, respectively.

Sample D is carbon-doped GaN structure grown on AlGaN/AlN buffer on a silicon (111) substrate. The dislocation density was estimated to be $2 \times 10^9$ cm$^{-2}$.

Sample E is 2-µm thick Fe-doped GaN (using Cp$_2$Fe as the precursor) grown on c-plane sapphire substrate with a dislocation density of $9 \times 10^8$ cm$^{-2}$.

Sample F is n-doped GaN grown on a silicon (111 substrate). The dislocation density is measured to be $2 \times 10^9$ cm$^{-2}$.

Sample G is non-polar (11-20) GaN grown using 3D-2D method on m-plane sapphire with a dislocation density[32] of $3 \times 10^{10}$ cm$^{-2}$.

Sample H is semi-polar (11-22) GaN grown on r-plane sapphire using silicon nitride interlayer and 3D-2D growth, which contain a typical dislocation density[40] of $4 \times 10^9$ cm$^{-2}$.

*Sample analysis*

Optical characterisation of the samples was performed in a lab-built confocal microscope using a 532-nm CW laser (Gem 532, Laser Quantum Ltd.) and a 710-nm CW laser (M2 SolsTiS Ti:Sapphire). Collection was done through a 0.9 NA air objective (TU Plan Fluor 100×, Nikon) and analysed either through a Spectrometer (SpectraPro Monochromator Acton SP2300, provided with a Pixis Camera 256, Princeton Instruments) or a pair of avalanche photodiodes (SPCM-AQR-14, Perkin Elmer) in a Hanbury-Brown and Twiss interferometer configuration. Samples were mounted on a XYZ positioning stage (P-611.3S, PI Nanocube). The AFM analysis was performed on a Dimension 3100 AFM in tapping mode. CL measurements were performed at room temperature in a Philips XL30s scanning electron microscope operating at 5 kV and equipped with a Gatan MonoCL4 system. TEM imaging was conducted in an FEI Tecnai Osiris operated at 200 kV. The sample was prepared by standard mechanical polishing followed by Ar$^+$ ion milling at 5 kV and polishing at 1 kV down to 0.1 kV.

The multi-microscopy analysis combining CL, AFM and PL measurements for the same area of the sample was done using a reference alignment grid patterned directly on the samples via electron beam lithography (EBL).

**Authors contribution**

All authors conceived the project. T. Z. fabricated the samples. M. N. and M. K. designed and conducted the measurements with assistance from C.B. Data analysis was conducted by M. N., M. K., T. Z. F. M., R. O and C.B. All authors discussed the results and co-wrote the manuscript.

**Acknowledgements**

The authors thank financial support from the Australian Research council (via DP180100077 and DE 1801100810). The Asian Office of Aerospace Research and Development grant FA2386-17-1-4064, the Office of Naval Research Global under grant number N62909-18-1-2025 are gratefully acknowledged. This research is supported by an Australian Government Research Training Program Scholarship.

**Figures**

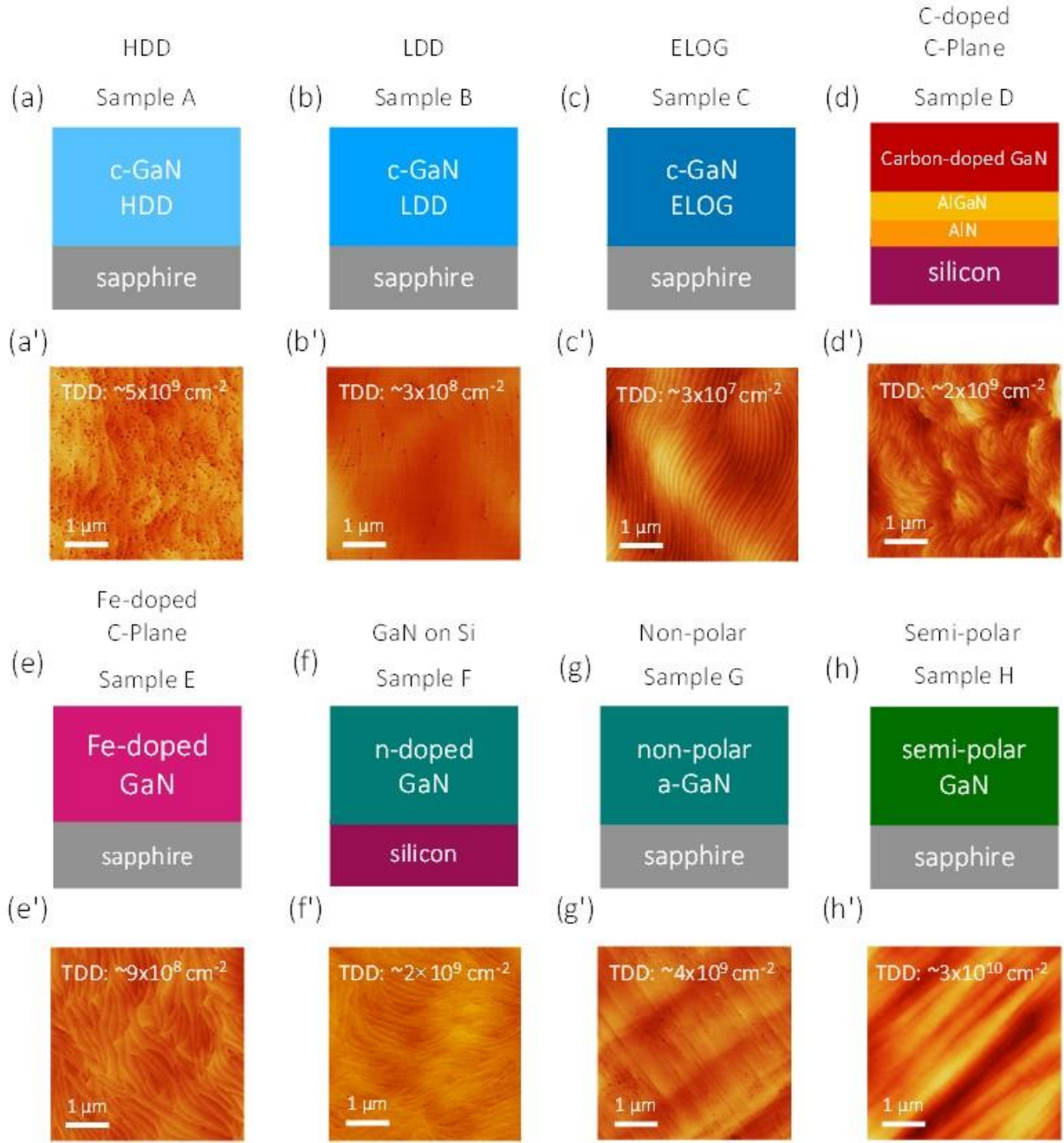

**Figure 1.** Schematic and topological characterisation of all GaN samples investigated. **a–h)** Schematics of c-plane HDD GaN (a), c-plane LDD GaN (b), c-plane ELOG GaN (c), C-doped GaN (d), Fe-doped

GaN (e), n-doped GaN grown on silicon (f), non-polar GaN (g) and semi-polar GaN (h). **a'–h')** Corresponding topological AFM scans for c-plane HDD GaN (a'), c-plane LDD GaN (b'), c-plane ELOG GaN (c'), C-doped GaN (d'), Fe-doped GaN (e'), n-doped GaN grown on silicon (f'), non-polar GaN (g') and semi-polar GaN (h'). The threading dislocation density (TDD) is noted for each topological scan.

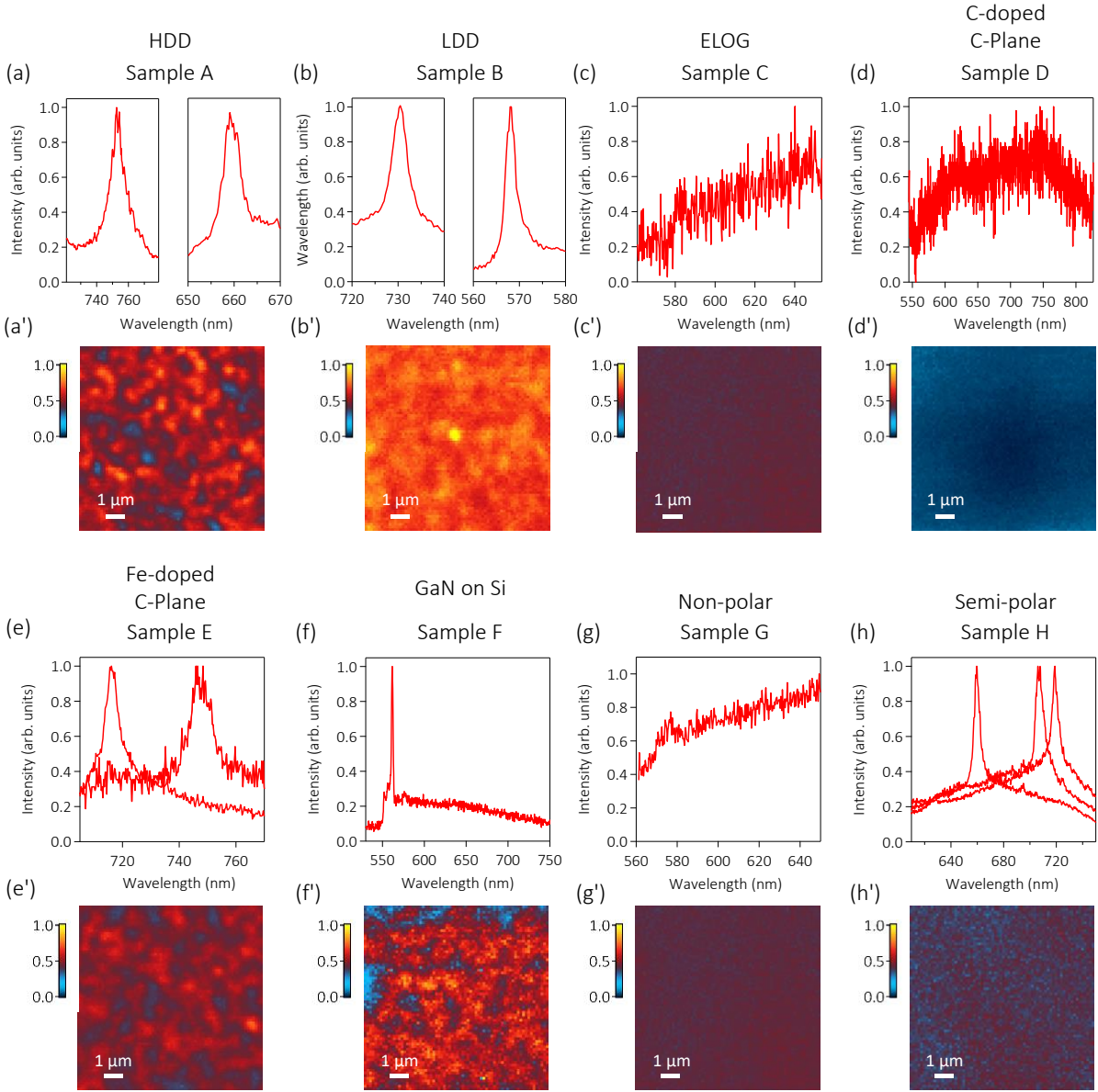

**Figure 2.** PL analysis of the emitters found in the GaN samples. **a–h)** Representative PL spectrum of the various samples. Sample A, B, E and H show distinct peaks associated to the PL of emitters, while samples C, D, F and G display a rather broad PL signal (the sharp peak in F is the Raman peak from the silicon substrate). **a'–h')** Corresponding 10 × 10 μm$^2$ confocal scans where the spectroscopy measurements were conducted. Specifically: HDD GaN (a, a'), LDD GaN (b, b'), ELOG GaN (c, c'), C-doped GaN (d, d'), Fe-doped GaN (e, e'), GaN on Si (f, f'), non-polar GaN (g, g') and semi-polar GaN (h, h'). All spectra were acquired upon excitation via a 532-nm CW laser.

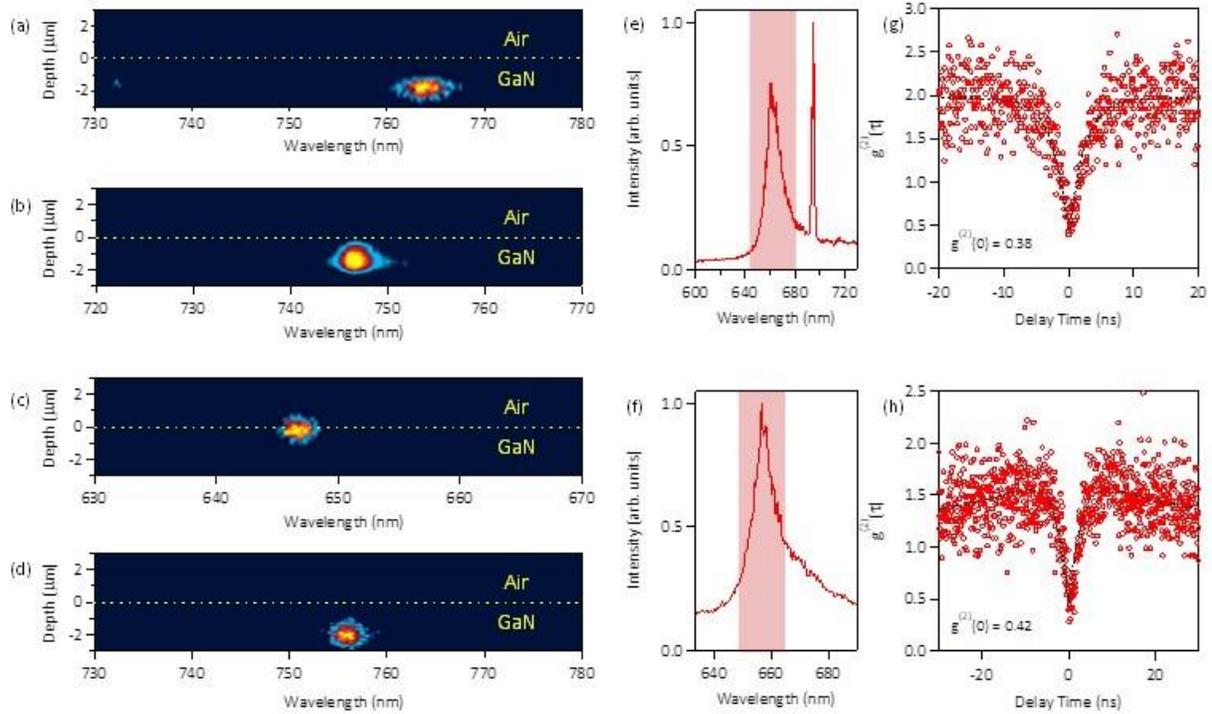

**Figure 3.** Optical analysis of emitters in GaN. **a–d)** Depth-resolved PL measurements of emitters in LDD (a), HDD (b), semi-polar (c), and (d) Fe-doped GaN. The yellow line indicates the boundary between the GaN surface and air. All depth scans were probed with 532 nm CW excitation. **e, f)** Spectrum of representative emitters in LDD GaN chosen for autocorrelation measurements. The red box indicates the bandwidth of the collected signal after filters were used to eliminate background emission. **g, h)** Corresponding second order autocorrelation measurements $g^{(2)}(\tau)$ of the emitters in (e) and (f) showing their quantum nature (i.e. $g^{(2)}(0) \leq 0.5$). Spectroscopy and autocorrelation measurements were performed with 532-nm CW excitation.

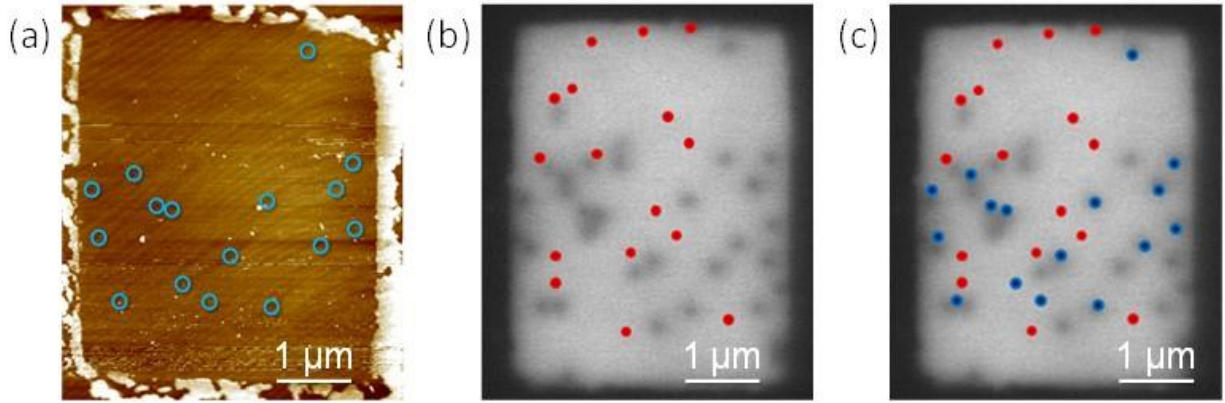

**Figure 4.** AFM, panchromatic CL, and PL measurements of LDD Sample B correlating the position of threading dislocations with the optical properties of GaN emitters. **a)** AFM image of the characterised grid showing the positions [blue circles] of the dislocations as they terminate at the surface. The yellow box indicates the borders of the grid. **b)** CL image of the same grid. The [red dots] indicate the position of photoluminescent emitters as characterised separately by confocal microscopy under 532-nm CW laser excitation (confocal scan not shown). The TDs appear as [black spots] in the CL image. **c)** Overlay of images (a) and (b) showing the correlation between surface features spatially characterised in AFM [blue dots] and CL [black spots], and emitters characterised in PL [red dots].

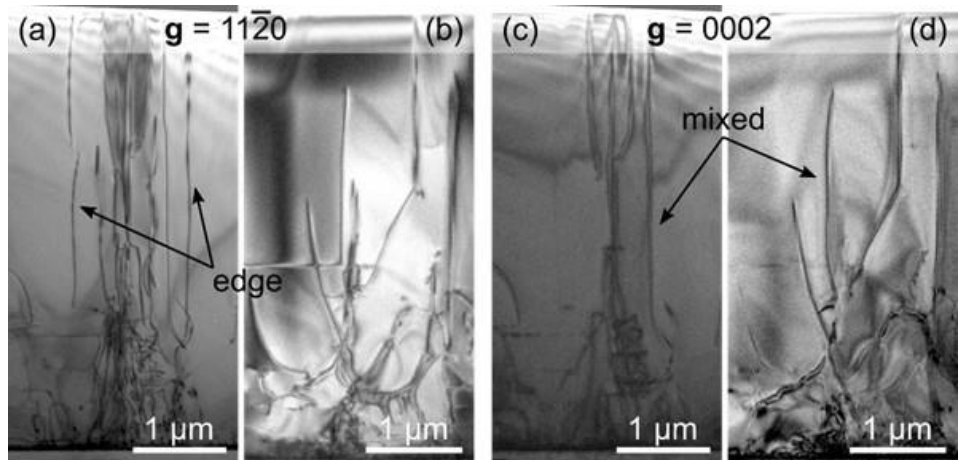

**Figure 5.** TEM analysis. **a–d)** Cross-sectional TEM images of Sample B (LDD), observed along the $\langle 1\bar{1}00 \rangle$ zone-axis and diffraction conditions $g = 11\bar{2}0$ (a, b) and $g = 0002$ (c, d). Edge-type dislocations (i.e. **b** = a) are visible under $g = 11\bar{2}0$ but not when $g = 0002$. Inversely, screw-type dislocations (i.e. **b** = c) are visible under $g = 0002$ but not when $g = 11\bar{2}0$—none could be seen in this image. Mixed-type dislocations (i.e. **b** = a+c) can be seen under both conditions.